\documentclass[%
 reprint,
 amsmath,amssymb,
 aps,
]{revtex4-1}

\usepackage{graphicx}
\usepackage{dcolumn}
\usepackage{bm}
\usepackage{subfig}

\begin{document}

\preprint{}

\title{ Understanding the Heavy Tailed Dynamics in Human Behavior}
\author{Gordon J Ross}
\affiliation{Department of Statistical Science, University College London, United Kingdom}
\email{gordon.ross@ucl.ac.uk}

\author{Tim Jones}
\affiliation{Department of Computer Science, University of Bristol, United Kingdom}
\email{tim.jones@bristol.ac.uk}
\date{\today}

\begin{abstract}
The recent availability of electronic datasets containing large volumes of communication data has made it possible to study  human behavior on a larger scale than ever before. From this, it has been discovered that across a  diverse range of data sets, the inter-event times between consecutive communication events obey heavy tailed power law dynamics. Explaining this has proved controversial, and two distinct hypotheses have emerged. The first holds that these power laws are fundamental, and arise from the mechanisms such as priority queuing that humans use to schedule tasks. The second holds that they are a statistical artefact which only occur in aggregated data when features such as circadian rhythms and burstiness are ignored. We use a large social media data set to test these hypotheses, and find that although models that incorporate circadian rhythms and burstiness do explain part of the observed heavy tails, there is residual unexplained heavy tail behavior which suggests  a more fundamental cause. Based on this, we develop a new quantitative model of human behavior which improves on existing approaches, and gives insight into the mechanisms underlying human interactions.
\end{abstract}

\pacs{Valid PACS appear here}
\maketitle



The prospect of finding quantitative models that can describe and predict human behavior has fascinated researchers for decades, partly because understanding such behavior is interesting in its own right, and partly because these models can have important practical uses in fields as diverse as network analysis \cite{Iribarren2009, Goncalves2008, Radicchi2009,Kitsak2012}, cyber security \cite{Neil2014, Scott2002}, and the analysis of terrorism \cite{Raghavan2014}. The increased availability of databases containing large volumes of electronic communication data such as phone call records, emails, and social media interactions, has now made it possible to study human behavior on a larger scale than ever before.

One area which has attracted a great deal of attention \cite{Barabasi2005,Oliveira2005, Malmgren2008, Wu2010, Vazquez2006, Jiang2013} is the modeling of human communication event times. For some person $i$, let $t_1, t_2, \ldots, t_{n_i}$ denote the times at which this person engages in a particular type of communication event, such as sending an email or making a phone call. The inter-event times $\tau_i$ are defined as the times that elapse between successive communication events, so that $\tau_i = t_{i+1} - t_i$. The most simple quantitative model of behavior assumes that the occurrence of these events obeys a simple Poisson process, so that the inter-event times are governed by a memoryless Exponential distribution $p(\tau_i) = \lambda e^{-\lambda \tau_i}$. If true, this would be an important finding since it would imply a high degree of regularity and predictability in human behavior. However, it has now been shown that many types of human activity seems to be fundamentally non-Poissonian, with the event time distribution $p(\tau_i)$ exhibiting heavy-tails that decrease at a slower than exponential rate  \cite{Barabasi2005, Vazquez2006, Vazquez2007, Jiang2013, Hong2009, Eckmann2004, Perello2008}.

The origin of this heavy-tailed behavior requires explanation, and two competing hypotheses have been put forwards. The first holds that heavy tails are fundamental to human behavior and arise through some variant of the priority queue mechanism \cite{Barabasi2005,Vazquez2006,Walraevens2012,Jo2012b}. The second holds that heavy tails are an emergent phenomena which are present only in aggregated data due to averaging over much simpler `local' behavior, such as circadian rhythms and burstiness.

An important quantitative model based on the latter hypothesis was recently introduced by \cite{Malmgren2008} and refined in \cite{Malmgren2009}, and argues that  heavy tails can be fully explained by a simple  model which takes into account empirical facts about how human  behave. This model assumes  that individuals can be in one of two states - active and inactive. The dynamic transition between these states incorporates both circadian rhythms and burstiness, and purports to fully explain heavy tailed behavior as an aggregation of simple Poisson processes. The reduction of complex human behavior down to interacting Poisson processes has great theoretical appeal, as well as being useful for practical purposes such as predicting the times of future events. As such, this model has been widely adopted \cite{Olson2010, Malmgren2009a, Anteneodo2010}.

Most previous comparisons of these two hypotheses have focused on email and mobile phone usage. In contrast, we here study the behavior of individuals on social media  which, as we will discuss in the next section, poses challenges for quantitive models for several reasons.  The networks resulting from the use of social media have become an important field of study within the complex systems literature, with interest focusing on both the behavior of individual users, and the interactions between multiple users \cite{Borge-Holthoefer2012, Borge-Holthoefer2012b, Zhao2010, Newman2003, Miritello2011}.

Our key finding is that that existing models such as \cite{Malmgren2008} which treat heavy-tails as emergent rather than fundamental are inadequate for describing social media behavior. Specifically, we find that a) the distribution of inter-event times in the active state often has substantially heavier tails than previously thought, so that heavy-tailed behavior seems to be somewhat fundamental rather than fully emergent, and b) human behavior seems to be more complex than suggested by a simple active vs inactive state  dichotomy, with most individuals seeming to have multiple different types of active state, roughly corresponding  to short intense bursts of events, and less intense bursts which have a longer duration.

Despite these empirical findings, the basic insight  that human behavior is fundamentally driven by circadian rhythms and bursty dynamics is found to still hold empirically, and remains important. As such, the truth seems to lie somewhere between the two competing hypotheses - circadian rhythms and burstiness explain some but not all of the observed heavy tails in human behavior, and this points towards a more fundamental cause. Motivated by our findings, we put forward a new quantitative model for human communications which allows for both heavy tailed behavior and multiple types of active states which corrects the defects in the existing approaches. Our model has important implications for both the understanding and prediction of how humans behave.

\vspace{-4mm}

\section{Data}

We study the behavior of users on Twitter, which is a popular online social media site. On Twitter, each user has a unique  personal page which they use to share messages (known as tweets) with their followers. These tweets can either be general broadcast messages aimed at their whole group of followers, or specific messages aimed at individual users. In the latter case, it is not unusual for two users to conduct an extended conversation with each other via tweets. It is this dual usage which makes Twitter an interesting area of study, since it  incorporates two very distinct types of user behavior. First, users may post isolated `broadcast' tweets intended to share news/events with their followers. Second, users may engage in direct conversation with other specific users which will result in them exchanging multiple tweets, often within a short period of time. This is in contrast to other types of communication data such as SMS messaging where most communications occur as part of a short conversation without any  broadcast behavior  \cite{Wu2010}.

Our data consists of the Twitter activity for $10,000$ randomly selected users from the users who were active between June 2009 and December 2009 with the constraint that each user in the sample must have sent at least 300 tweets during that period, with a minimum 3 month gap between their first and last tweet,   For each user, we have access to the times at which all their tweets were made over the observation period.   Suppose that user $i$ made $N_i$ tweets during the 7 month observation period, and let the times of these tweets be denoted by $t_{i,1}, t_{i,2},\ldots,t_{i,3}$. The inter-event times for this user are then defined as $\tau_{i,j} = t_{i,j+1} - t_{i,j}$.

\begin{figure}[]
  \centering
  \subfloat[]{\label{fig:intro11}\includegraphics[width=0.25\textwidth]{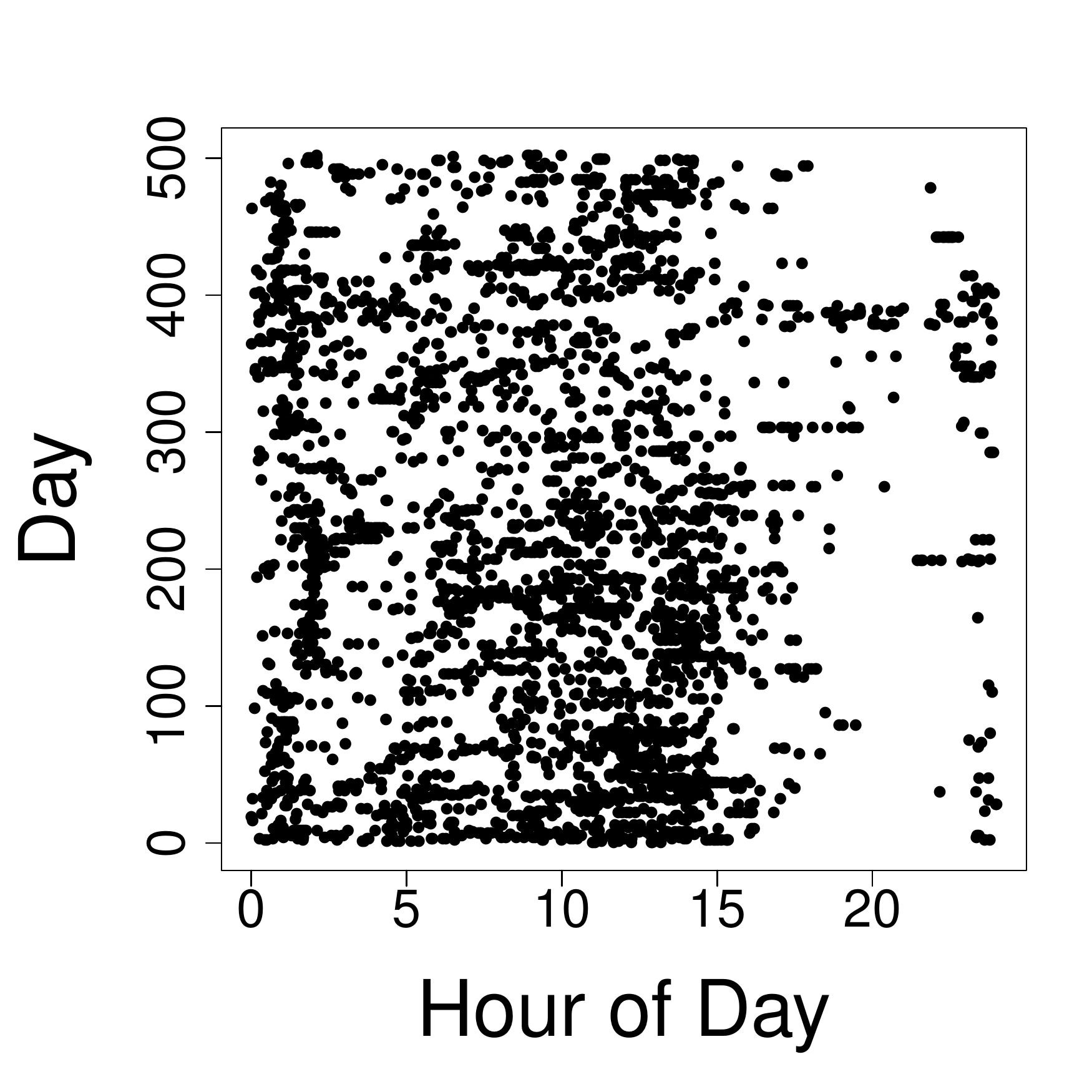}}
  \subfloat[]{\label{fig:intro12}\includegraphics[width=0.25\textwidth]{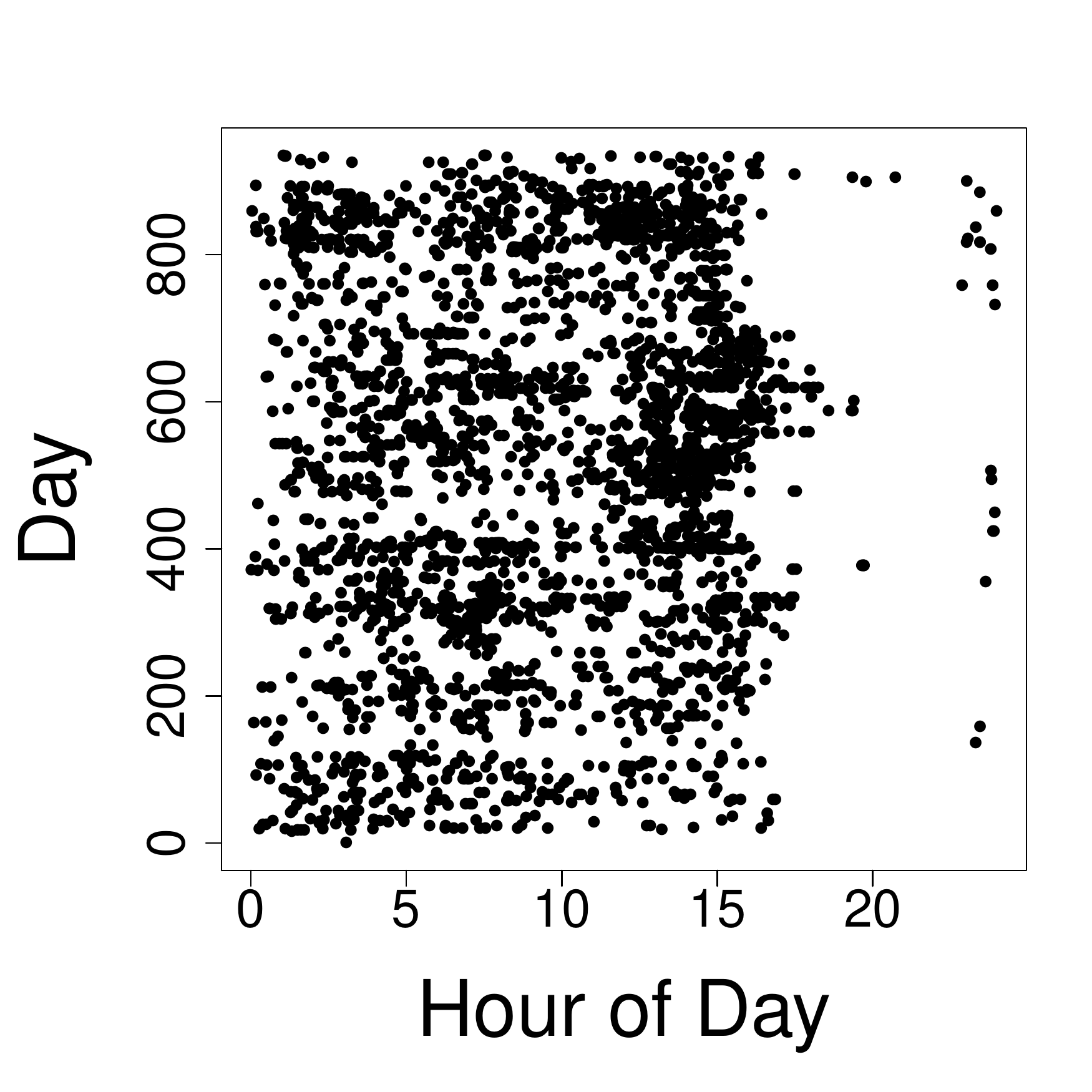}} \\
  \subfloat[]{\label{fig:intro13}\includegraphics[width=0.25\textwidth]{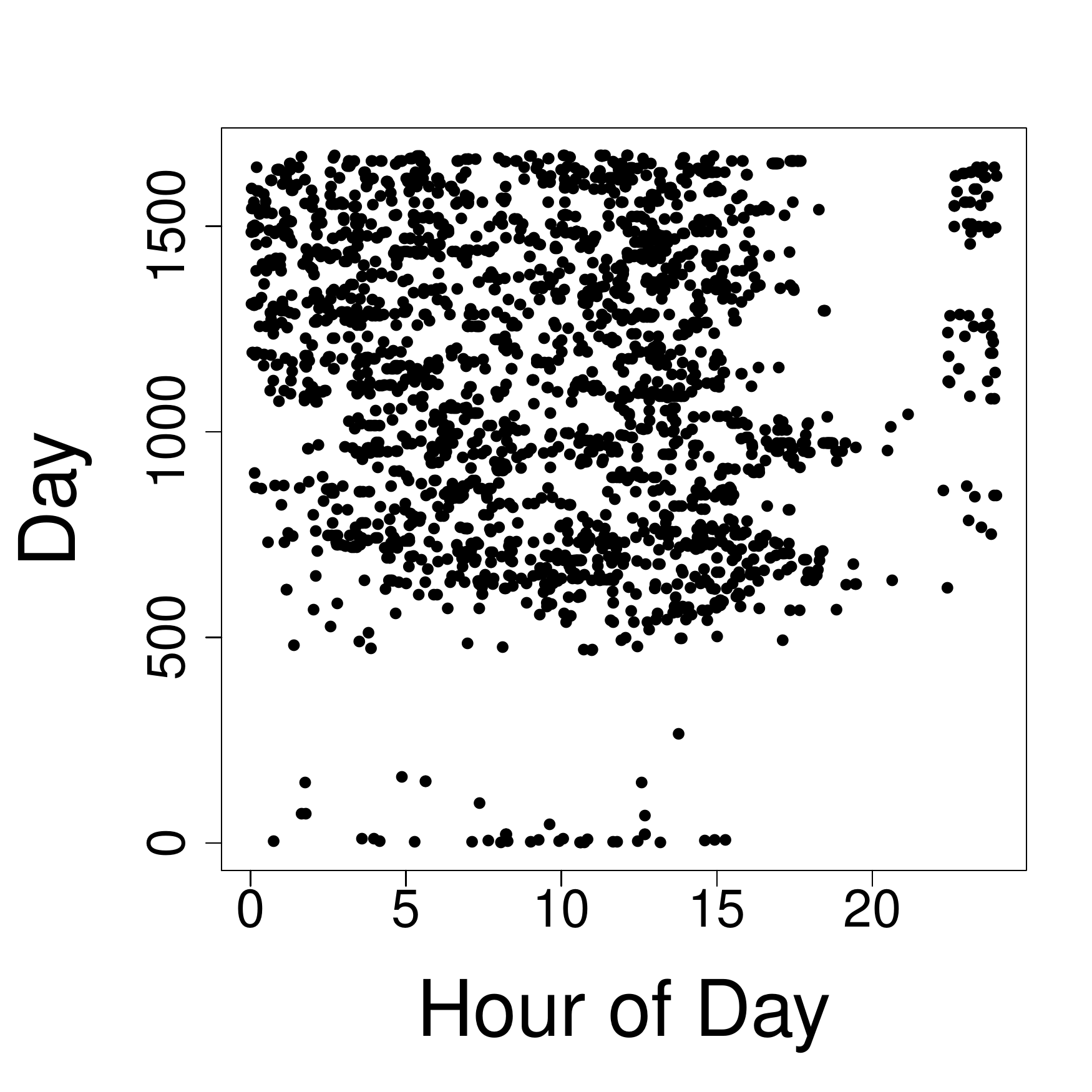}}
  \subfloat[]{\label{fig:intro14}\includegraphics[width=0.25\textwidth]{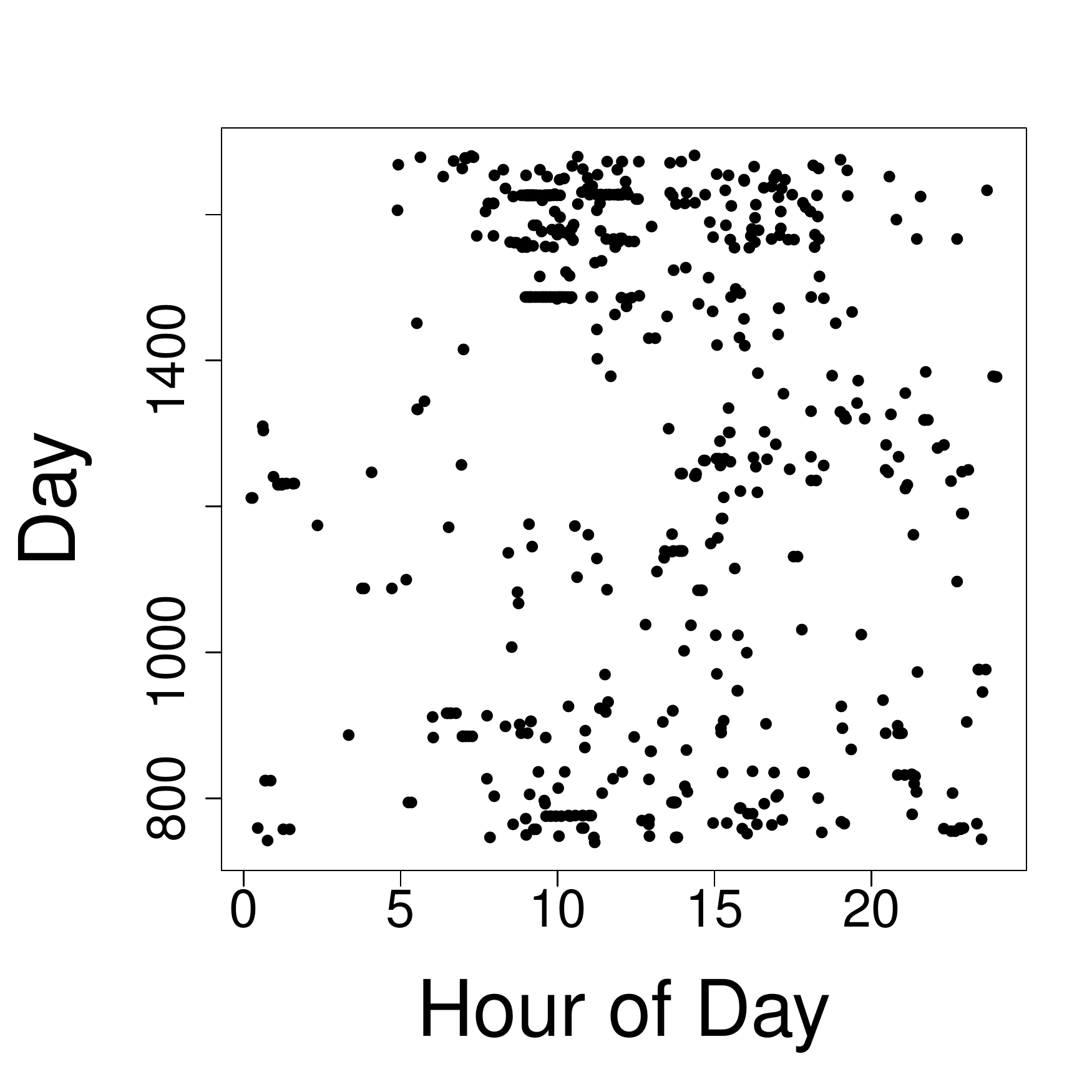}}
\caption{Example data for four Twitter users. Each plot represents the event sequence of one user, where each row is a single day and the days are stacked on top of each other. Dots represent the times at which each user made tweets.}
  \label{fig:userdata}
\end{figure}

\begin{figure}[]
  \centering
  \subfloat[]{\label{fig:intro21}\includegraphics[width=0.25\textwidth]{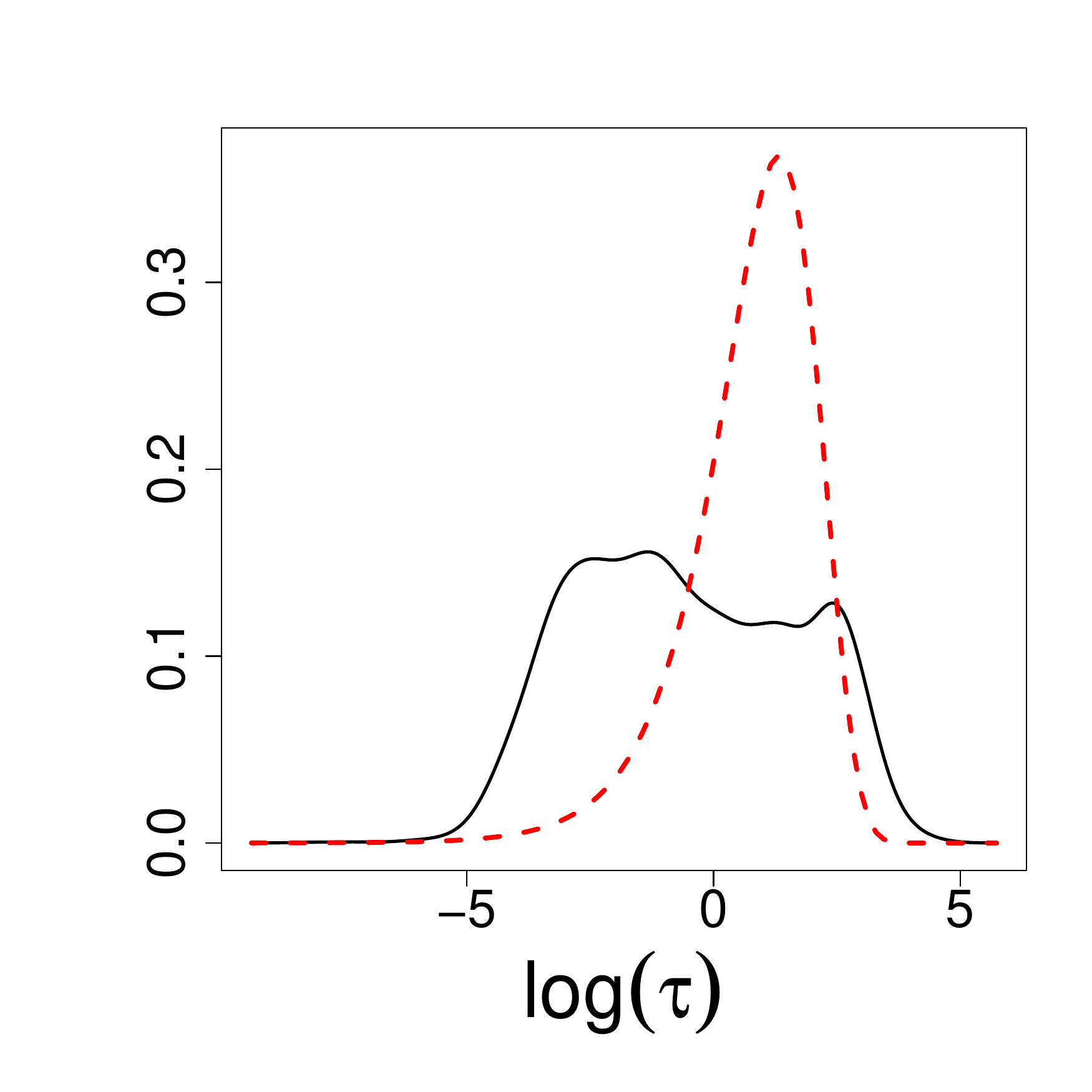}}
  \subfloat[]{\label{fig:intro22}\includegraphics[width=0.25\textwidth]{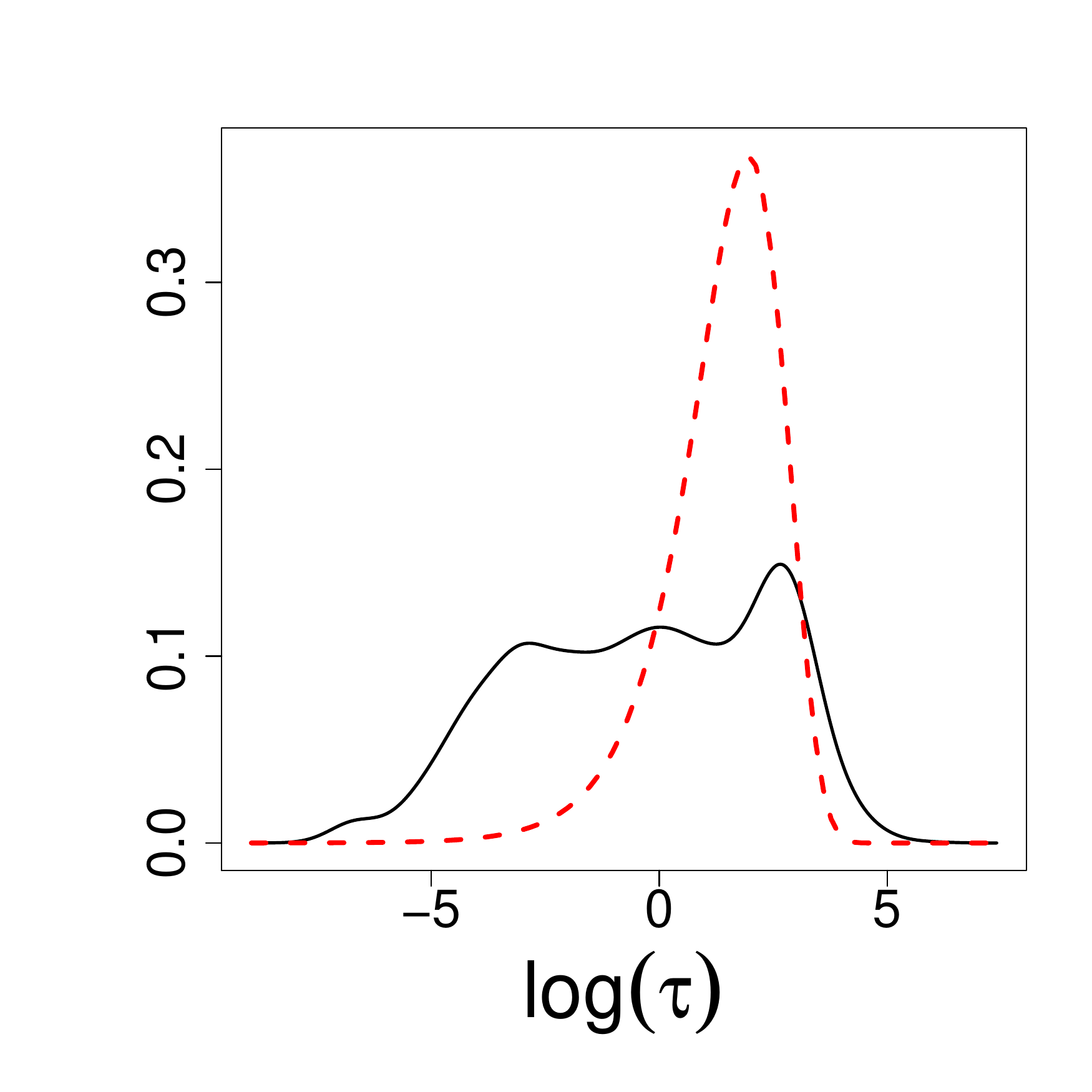}} \\
  \subfloat[]{\label{fig:intro23}\includegraphics[width=0.25\textwidth]{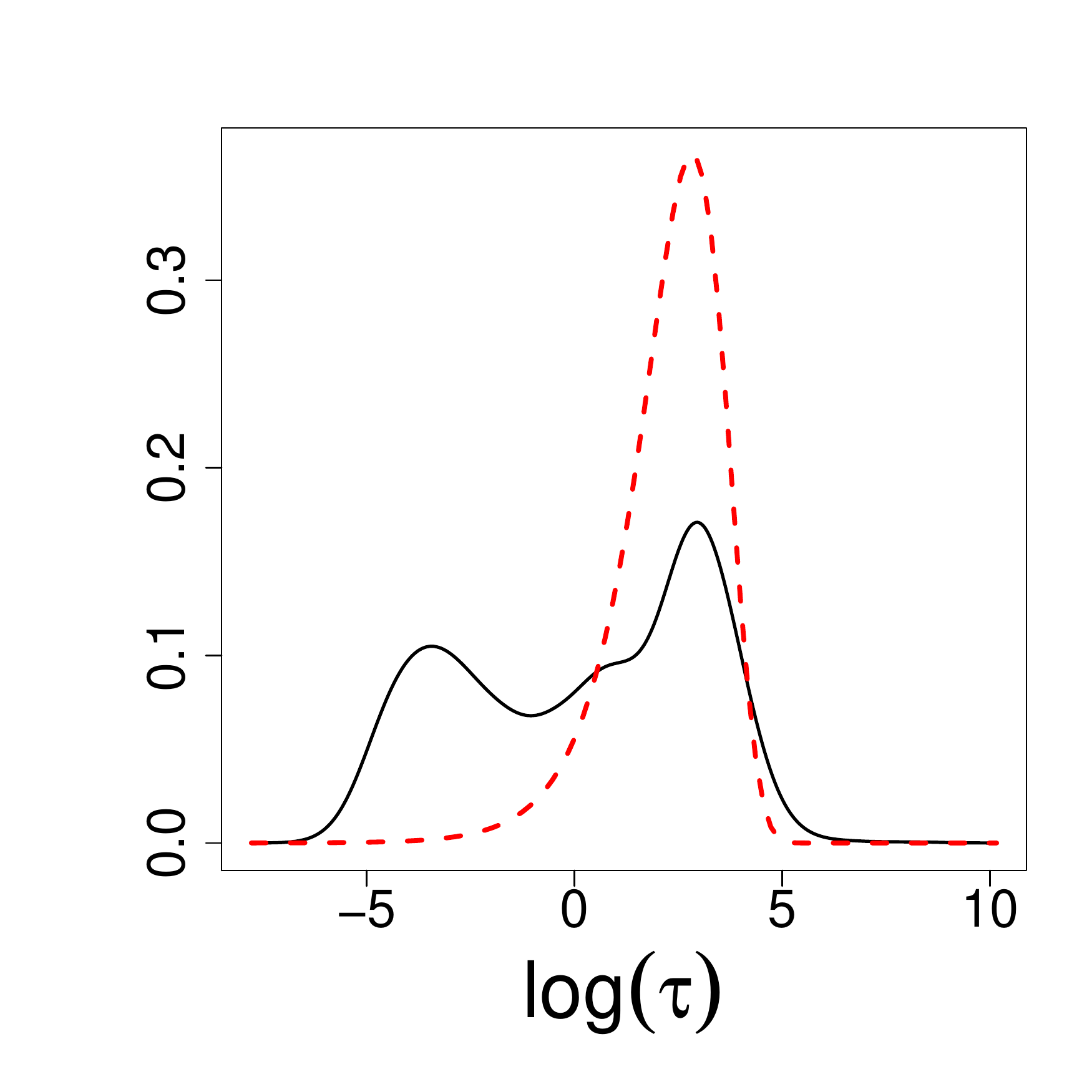}}
  \subfloat[]{\label{fig:intro24}\includegraphics[width=0.25\textwidth]{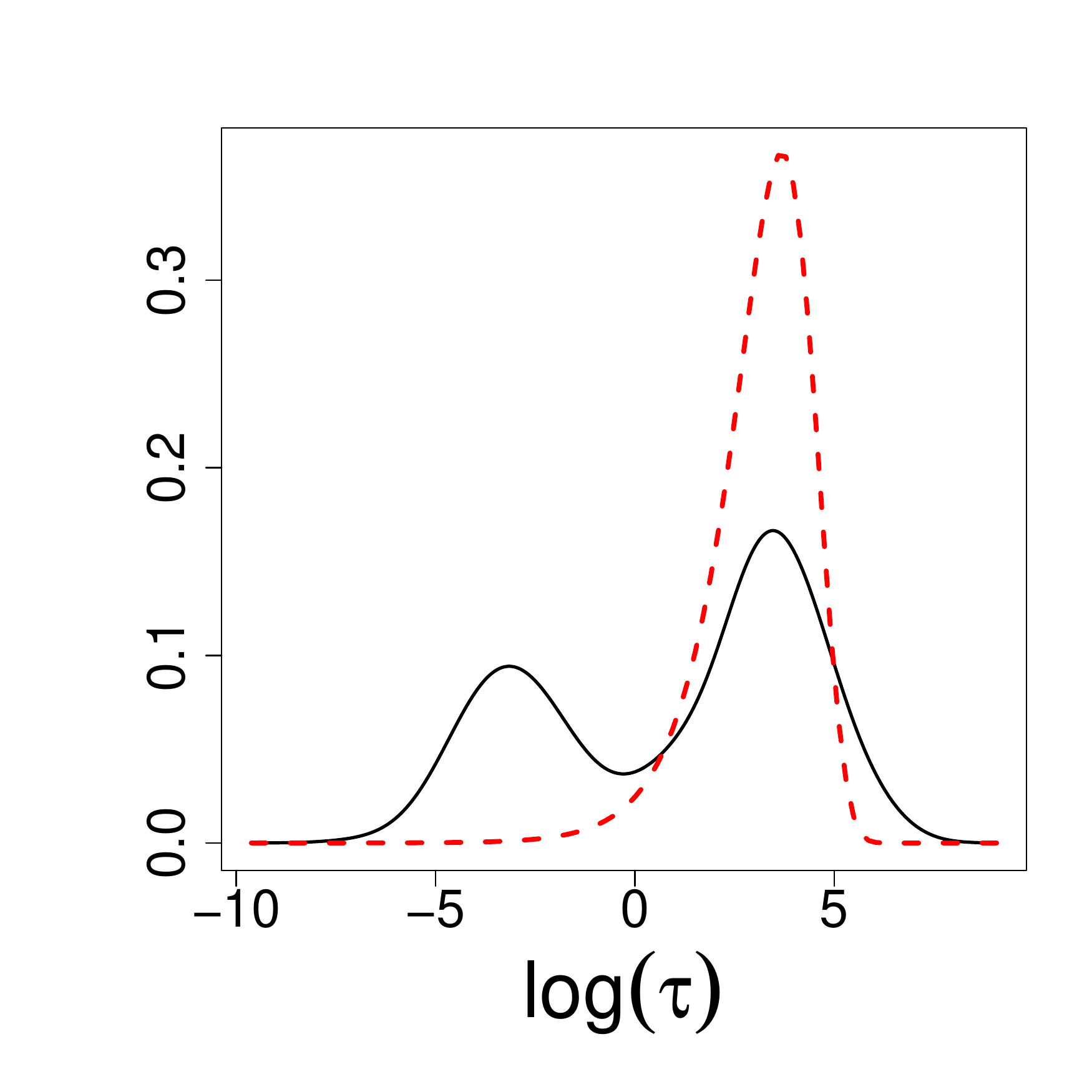}}
\caption{(Color online) Empirical density (black line) of the logarithm of the inter-event times for each of the four users, on a logarithmic scale. The dotted red line denotes the best fitting log-exponential distribution (note: the logarithmic transformation has been used to aid interpretation, and all logarithms are in base $e$)}
  \label{fig:twostatedensity}
\end{figure}


Fig. 1 shows example data for four particular Twitter users, where each plot corresponds to one  user, and each row represents one day. The different days in the sample period are stacked on top each other, with the bottom row corresponding to day 1. The times at which tweets were made are shown by dots. It can be seen that there are clear circadian rhythms, with all four users unsurprisingly having a consistent (but time-zone dependent) period of around 8 hours where they are usually inactive, which probably corresponds to the time when they are asleep. This data also shows strong evidence of bursty behavior, with tweets tending to occur in clusters, so that a person will regularly go several hours without tweeting and then send multiple tweets in a short period of time. These two features (circadian rhythms and burstiness) are also observed in the communication records of most of the other users in our sample, suggesting that at this high level, Twitter usage shares the same general patterns that have previously been noted in telephone, email, and SMS communications. \cite{Karsai2012,Wu2010,Jo2012}.

In Fig. \ref{fig:twostatedensity}, we plot the empirical density of the logarithm of the inter-event times $\tau$ for each of these four users. On each plot, we have superimposed the best fitting Exponential distribution (on a logarithmic scale, to aid visual interpretation). It can be seen visually that the Exponential distribution gives a very poor fit to the inter-event times, showing that the behavior clearly does not follow a regular Poisson process. For each of these four users, the empirical inter-event time densities have  far heavier tails than predicted under the Poisson assumption, and also tend to be bimodal. The same pattern is found for the vast majority of the other 10,000 users in our sample. In the next section we will describe how this type of non-Poissonian behavior can be modeled.


\section{The Simple Poissonian Burst Model and its Limitations}

We first recap the model introduced \cite{Malmgren2008} which incorporates both circadian rhythms and Poissonian bursts of events, and represents the current state of the art for modeling human behavior. We will refer to this as the Simple Poissonian Burst Model (SPB). Next, we will demonstrate that the SPB does not capture important features of social media behavior, and introduce a new model which is more appropriate.

Under the SPB, individual people have two states, inactive and active. Conceptually, the inactive state represents the periods when the person is not sending messages, while the active state represents the periods when they are. Most people are assumed to spend the majority of their time in the inactive state, but every so often, people switch to the active state, produce a burst of events which follows a homogenous Poisson process, and then revert to being inactive. As such, each day is partitioned into a sequence of active and inactive intervals, and events are clustered into active bursts.

The times when each individual switches to the active state are modeled by a time-inhomogenous Poisson process with an intensity function $\lambda_0(\cdot)$. This time-inhomogeneity allows for circadian rhythms, with $\lambda_0(\cdot)$ being high during the times of day when the person is most likely to be active, and low during the times when they are inactive (such as during the night). When this Poisson process produces an event, a single communication event is observed and the individual switches to the active state with probability $p$. While in the active state, multiple events are expected to occur due to the bursty nature of human behavior. Specifically, the events within each active burst are assumed to follow a homogenous Poisson process with intensity $\lambda_A$, where $\lambda_A >> \max \lambda_0(\cdot)$. After a certain number of  events have occurred, the burst is finished and the person reverts back to the inactive state, until the next burst. The precise number of events occurring within a burst is assumed to follow a Geometric distribution, with parameter $\theta$.
 
As noted in the previous section, the appeal of this model is that the events within each burst obey a Poisson process, implying that human behavior is very regular within each active session, and that heavy-tailed properties only appear in aggregate data due to  averaging over active bursts, rather than being fundamental to human behavior.

Fitting this model to the communication data associated with a particular person requires estimation of the person-specific parameters $\Theta = (\lambda_0(\cdot), p, \lambda_A, \theta)$.  To ease estimation, it is useful to rephrase the SPB model as a two-state Hidden Markov Model (HMM) \cite{Malmgren2009}. In this formulation, a hidden (unobserved) latent variable $s_i$ is associated with each inter-event time $\tau_i$. If $s_i = 0$, then $\tau_i$ was generated while in the inactive state, and if $\tau_i = 1$  then $\tau_i$ was generated while in the active state. The distributions of inter-event times under this model are then: 

\vspace{-2mm}

$$p(\tau_i | s_i, t_i, \Theta) \sim  \left\{ \begin{array}{ll}
 \mathrm{Exponential}(\tau_i | \Lambda_{t_i}^{t_i+\tau_i}) &\mbox{if $s_i=0$} \\ 
  \mathrm{Exponential}(\tau_i | \lambda_A) &\mbox{if $s_i= 1$,}
       \end{array} \right.
$$
where $\Lambda_a^b = \int_{a}^{b} \lambda_0(u) du$.  All the parameters associated with this model can be estimated using standard techniques for Hidden Markov Models, and details of this are provided in the Appendix. Crucially for our analysis, fitting this model to data also produces, for each $\tau_i$, an estimate of the associated value of $s_i$ -- i.e. whether each $\tau_i$ was generated during an inactive or active state.

This completes the description of the SPB model. The key question is now whether this model accurately describes human behavior -- i.e. are the assumptions embedded in the SPB model accurate? To answer this in the case of social media data, we fit the SPB model to the Twitter event sequence associated with each individual in our sample, estimating a different set of parameters for each user. The key assumption of the SPB model is that each active burst is described by a homogenous Poisson process, and so the inter-event times between events in the active state should follow an Exponential distribution with parameter $\lambda_A$, where $\lambda_A$ differs for each individual. This assumption can easily be tested since as noted above, the output of fitting the HMM to data includes an estimate of which observations were produced during active bursts. Therefore, the empirical active state inter-event time distribution can be constructed, and compared to the best fitting Exponential distribution. Fig. 3 shows these empirical active state distributions for the four Twitter users considered earlier, with the best fitting Exponential distribution superimposed in red. It can be seen that the fit is clearly superior to that previously observed in Fig. 2 when we considered only the aggregated inter-event times without breaking them down into active vs inactive states. As such, the SPB seems to explain much of the heavy-tailed behavior.  But despite this clear improvement, the Exponential distribution still does not seem to give an adequate fit for any of these four individual , and fails quite drastically for the 2nd and 3rd users in particular.

The same finding holds for the majority of the 10,000 other Twitter users in our sample. In general, the observed active state inter-event time distribution can differ quite substantially from the Exponential distribution predicted under the SPB model. Specifically, it seems that the empirical inter-event time distributions typically have heavier tails than the Exponential distribution allows. This suggests that the heavy tails observed in aggregated communication data are not explained by a model which assumes purely Poissonian local dynamics -- the heavy tails appear to be partially fundamental, rather than emergent.



\begin{figure}[]
  \centering
  \subfloat[]{\label{fig:intro51}\includegraphics[width=0.25\textwidth]{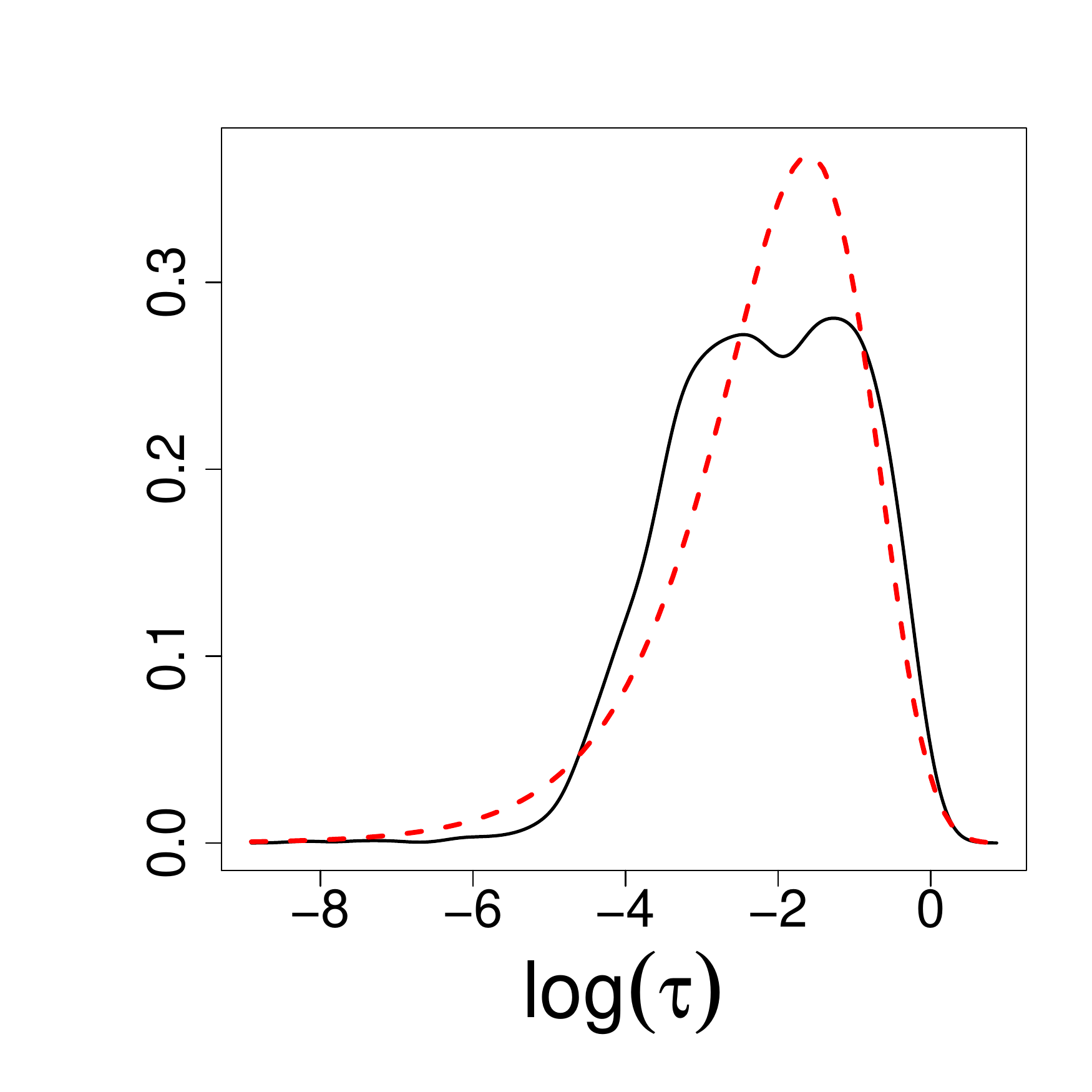}}
  \subfloat[]{\label{fig:intro52}\includegraphics[width=0.25\textwidth]{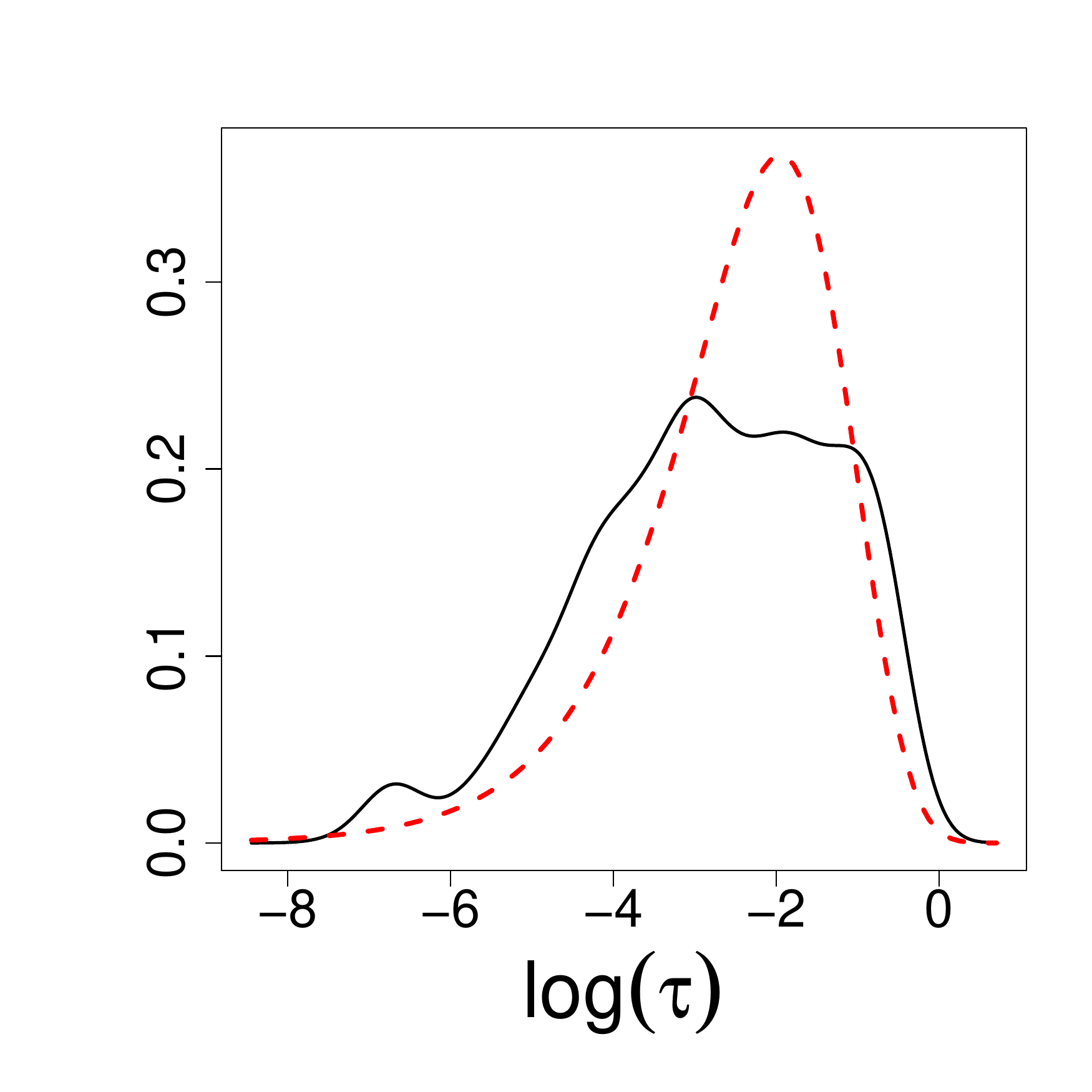}} \\
  \subfloat[]{\label{fig:intro53}\includegraphics[width=0.25\textwidth]{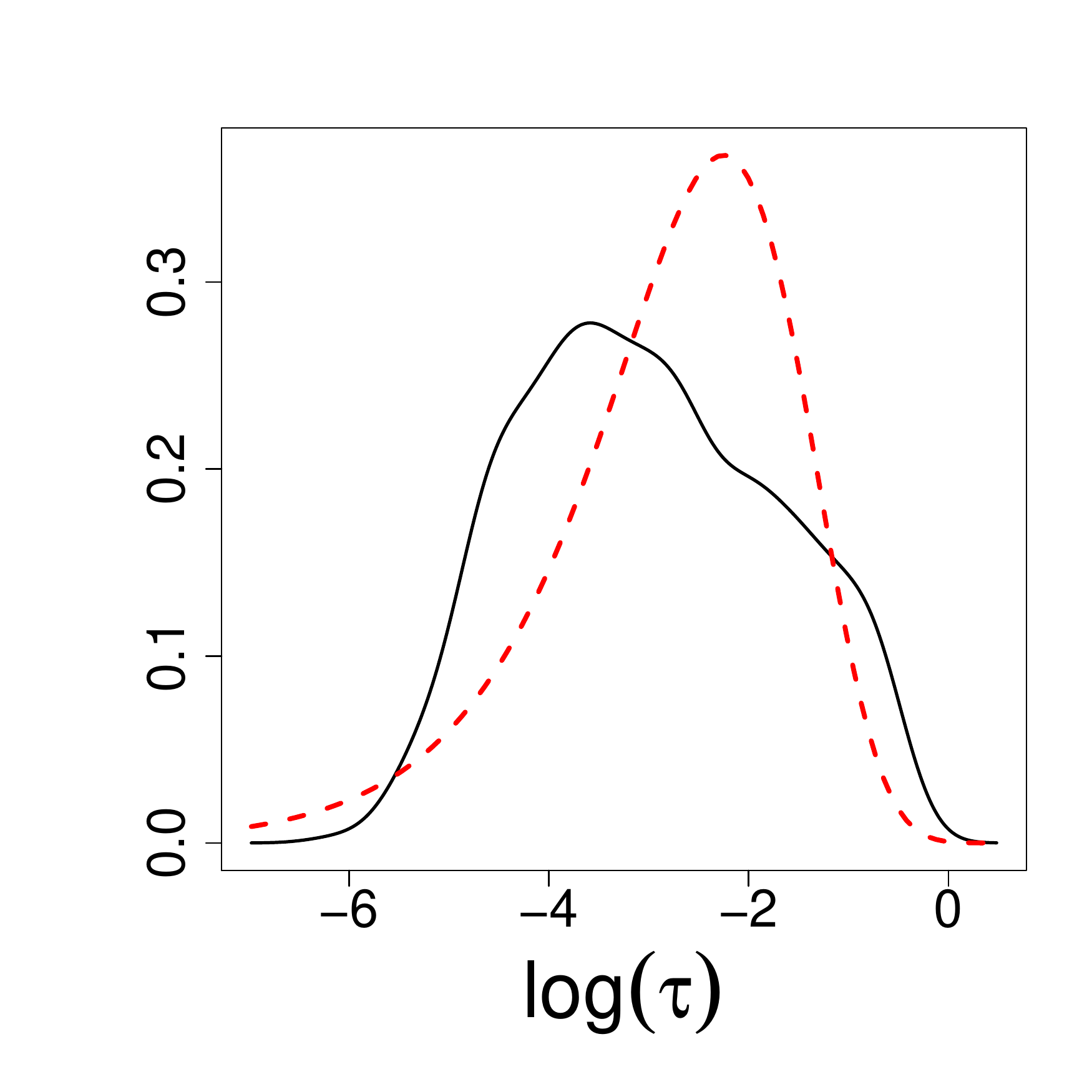}}
  \subfloat[]{\label{fig:intro54}\includegraphics[width=0.25\textwidth]{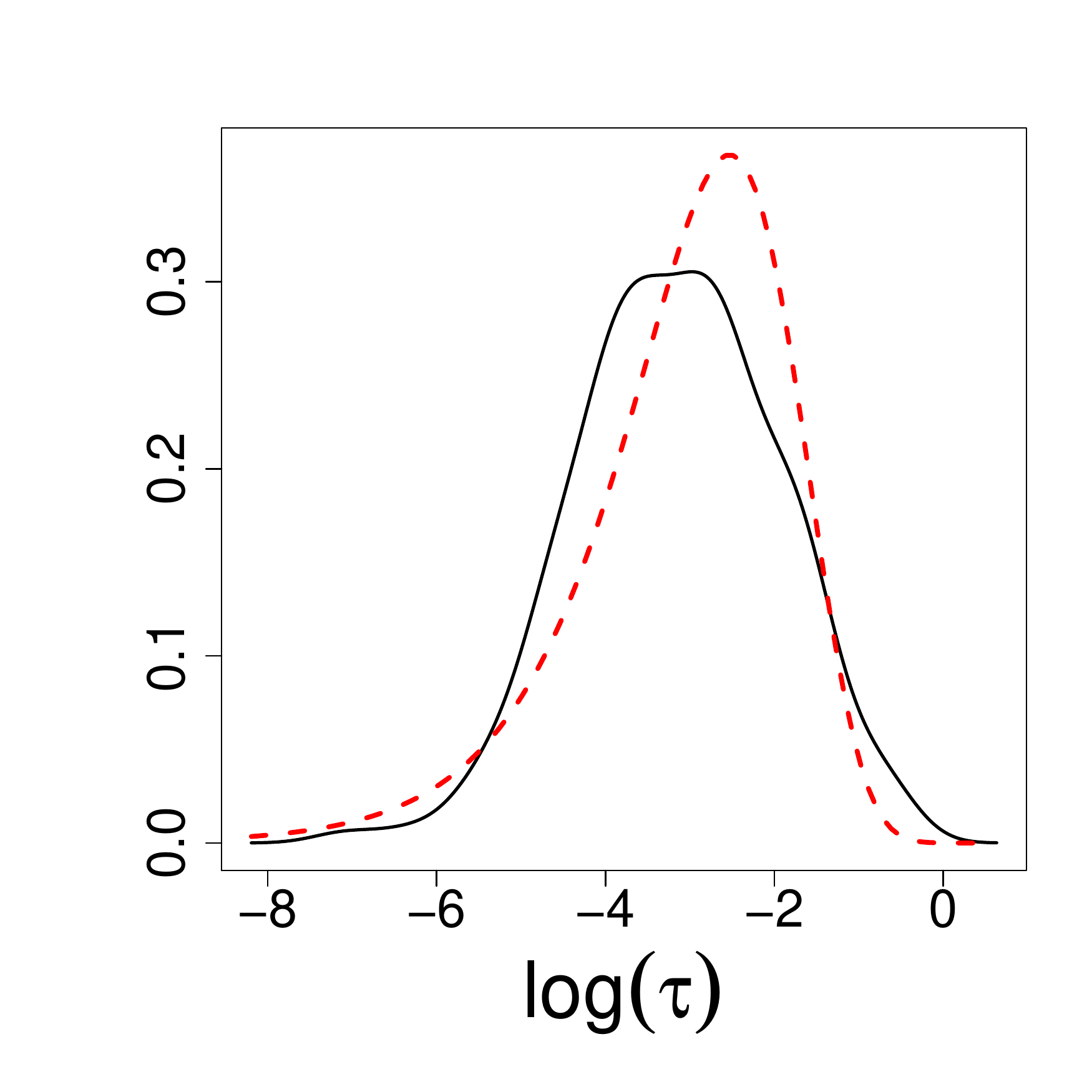}}
\caption{(Color online) Empirical density (black line) of the logarithm of the active state inter-event times for each of the four users, after fitting the simple Poissonian burst model. The dotted red line denotes the best fitting log-exponential distribution (note: the logarithmic transformation has been used to aid interpretation)}
  \label{fig:twostatedensity2}
\end{figure}

\vspace{-2mm}

\section{Incorporating Heavy Tailed Inter-Event Distributions}

Based on the evidence in the previous section, we propose a new model for human behavior which explicitly allows for heavy tailed within-burst inter-event times and can hence replicate the findings we observed. Specifically, we consider replacing the Exponential distribution used in the SPB active state with a more flexible (yet still theoretically justifiable) distribution which can adapt to heavy/light tails as required.  Our model is hence:

\vspace{-2mm}

$$p(\tau_i | s_i, t_i, \Theta) \sim  \left\{ \begin{array}{ll}
 \mathrm{Exponential}(\tau_i | \Lambda_{t_i}^{t_i+\tau_i}) &\mbox{if $s_i=0$} \\ 
  g(\tau_i | \gamma) &\mbox{if $s_i=1$,}
       \end{array} \right.
$$
where $g(\cdot)$ is a suitably chosen inter-event time distribution with parameter vector $\gamma$. Note that we have retained the Exponential distribution in the inactive state since the inhomogenous Poisson process governing this state is already capable of producing heavy-tailed behaviour due to the time-varying $\Lambda$ parameter.

Although there are many possible forms that could be chosen for $g(\cdot)$, we will specifically consider both the Lognormal and Weibull distributions since there is some preliminary evidence from the existing literature that suggests they are well suited to modeling human behavior. The Lognormal distribution is parameterized by a mean and variance parameter $\gamma = (\mu, \sigma^2)$ in log-space, and is well-known to produce tails which are reminiscent of power laws \cite{Clauset2009, Stouffer2006}. It is also justifiable  as a model for human behavior  from a theoretical standpoint, due  to the central limit theorem from elementary statistics, which implies that it is the limiting distribution of any random variable which arises from multiplicative fluctuations in more elementary processes. The inter-event times $\tau_i$ have a Lognormal distribution $g(\tau_i |\gamma) = LN(\tau_i | \mu, \sigma^2)$ if


$$g(\log(\tau_i)|\mu,\sigma^2) \sim \frac{1}{\sqrt{2 \pi \sigma^2}} \exp\left(-\frac{(\log(\tau_i)-\mu)^2}{\sigma^2}\right)$$

The Weibull distribution also allows for heavy tailed behavior that can incorporate power-laws, and was found by \cite{Jiang2013} to be a good model for aggregated inter-event times in SMS data, when burstiness and circadian rhythms are not modeled explicitly. The inter-event times follow a Weibull distribution with parameter vector $\gamma = (\alpha, \beta)$ if:

$$g(\tau_i | \alpha, \beta) \sim \frac{\alpha}{\beta} \left( \frac{\tau_i}{\beta} \right)^{\alpha-1} \exp \left( -\frac{\tau_i^\alpha}{\beta^\alpha} \right)$$

We fit a two-state Hidden Markov Models to all 10,000 Twitter users using each of these two distributions for the active state inter-event times. Combined with the SPB in the above section, this means that we are fitting three separate models to each user, corresponding to Exponential, Lognormal, and Weibull active state inter-event distributions. The interesting question is then which of these models is most suitable for describing social media usage. To answer this, we perform model selection in a  standard manner, and define the best fitting model for each individual as the model which maximizes the log likelihood of the data, penalized by a quantity which prevents over fitting. This penalty is necessary since both the Lognormal and Weibull models have an additional parameter compared to the Exponential distribution. We use the standard Akaike Information Criterion (AIC) penalty \cite{Akaike1974} which penalizes based on the number of parameters in the model -- note that other criterion such as the Bayesian Information Criterion could also have been used but were deemed less appropriate since they penalize based on the total number of observations, while in our context only the observations in the active state contribute to the difference between models. Using the AIC criterion means that for each individual $i$ we choose the model which maximizes:

\vspace{-2mm}

$$\log L(\tau_{i,1},\ldots,\tau_{i,N_i} | M_j) - k_j $$
where $L(\cdot)$ is the likelihood of the observed data, $M_j$ is a model indicator variable where $j \in \{1,2,3\}$ corresponds to each of the three models under consideration, and $k_j$ denotes the number of parameters in model $j$. For each of the the three models, there is a single parameter which correspond to the probability of starting in the active and inactive state at the beginning of the observed period, along with the $\theta$ parameter governing the Geometric distribution which defines the number of events in each active burst, and the $p$ parameter governing the probability of transitioning to the active state when an inactive event occurs. The Exponential model has an additional 1 parameter, while the Lognormal and Weibull models have an additional 2 parameters. As such, we have $k=4$ for the Exponential model, and $k=5$ for the Lognormal and Weibull models. Note that there are also additional parameters corresponding to the parameterization of $\lambda_0(\cdot)$ as described in the Supplementary Material, but since this parameterization is the same across all models it can be ignored since its contribution to the penalty term is invariant.

\begin{figure}[]
  \centering
  \subfloat[]{\label{fig:intro31}\includegraphics[width=0.25\textwidth]{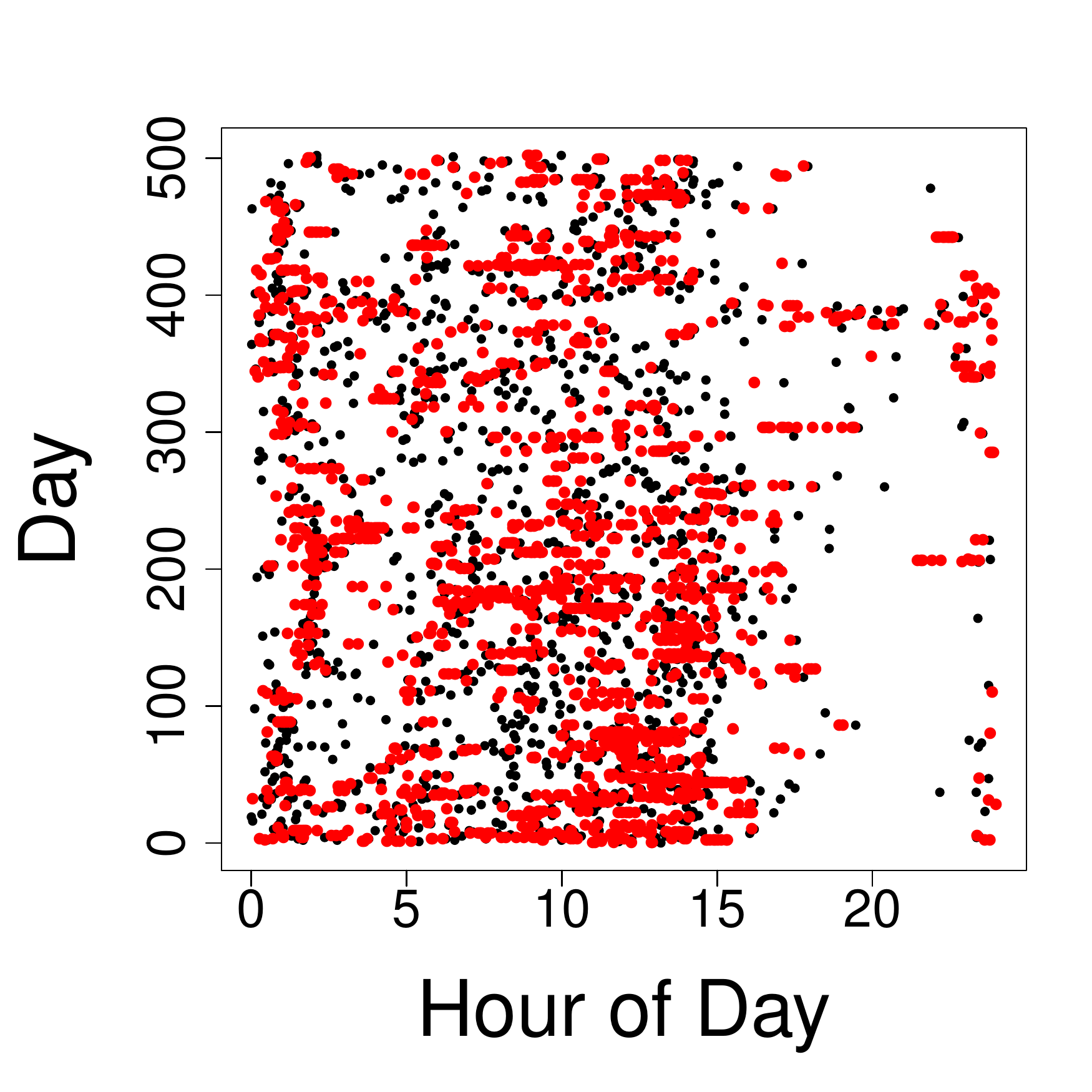}}
  \subfloat[]{\label{fig:intro32}\includegraphics[width=0.25\textwidth]{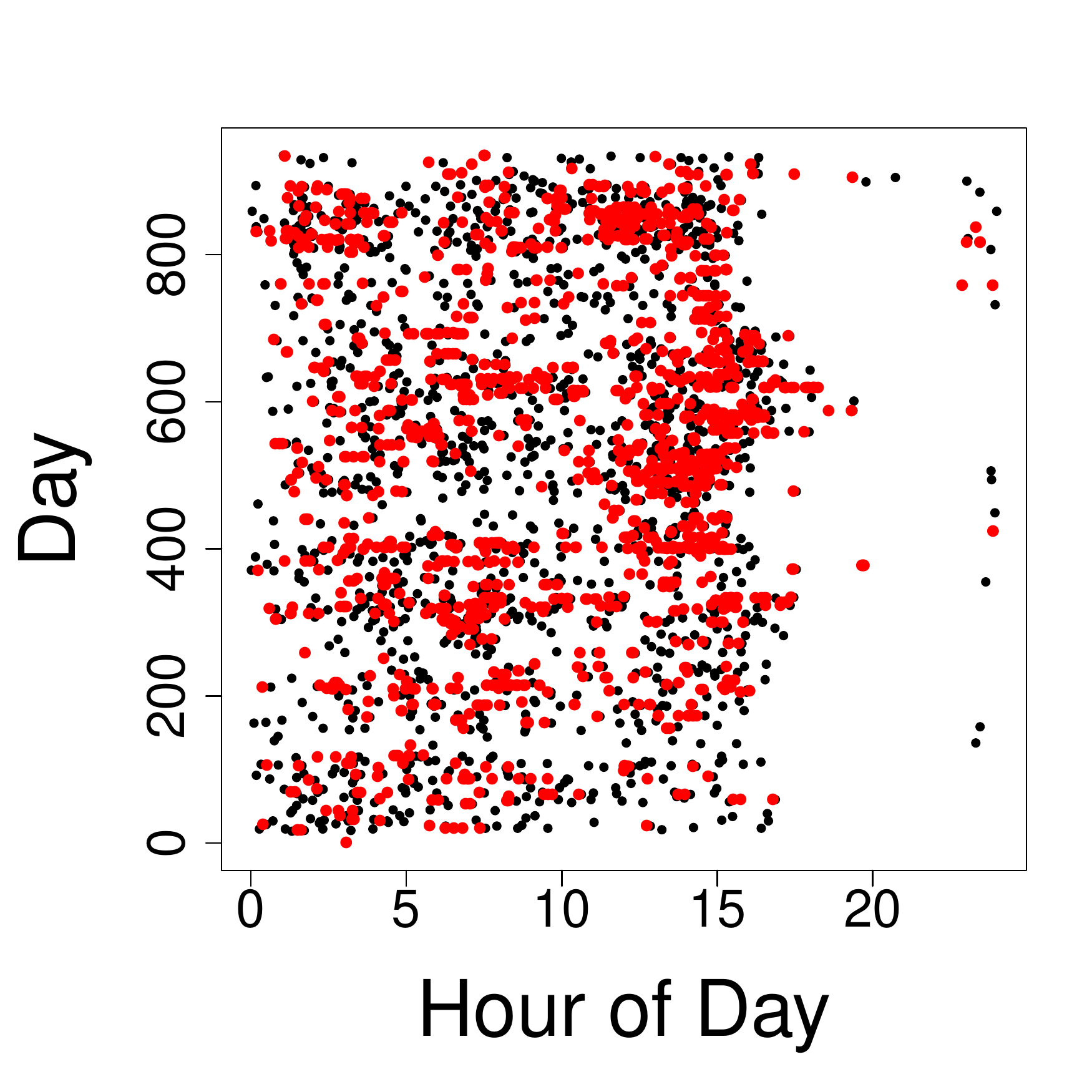}} \\
  \subfloat[]{\label{fig:intro33}\includegraphics[width=0.25\textwidth]{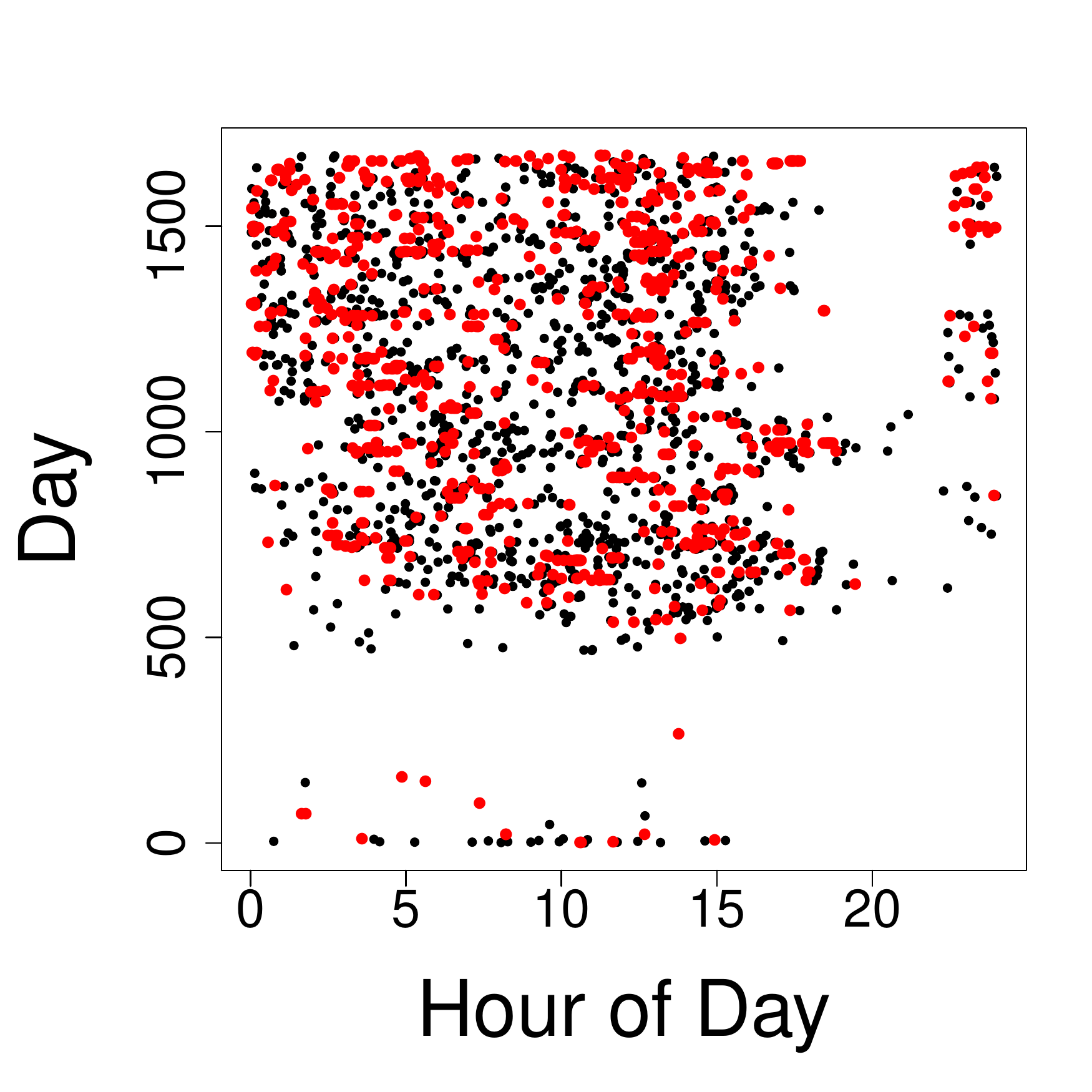}}
  \subfloat[]{\label{fig:intro34}\includegraphics[width=0.25\textwidth]{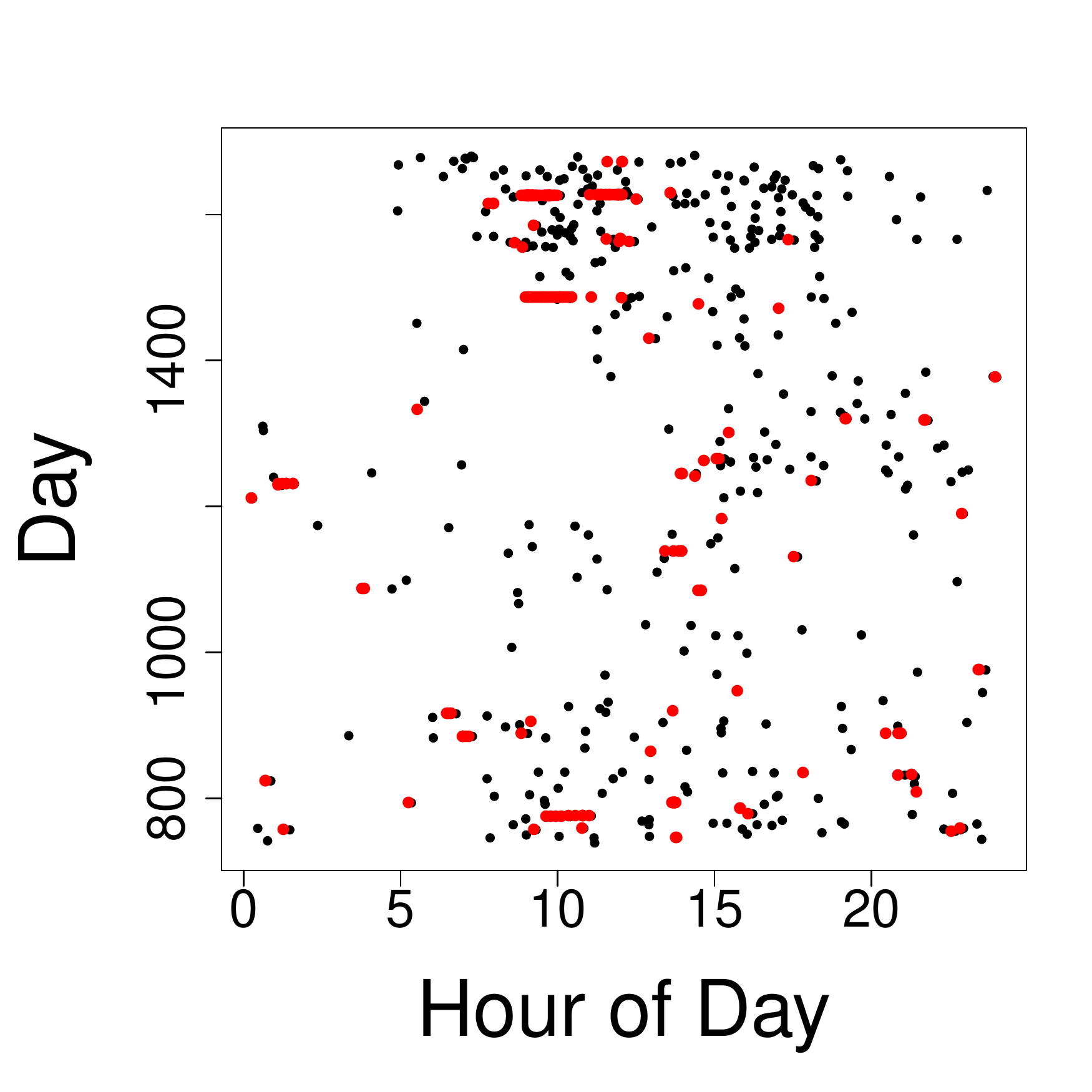}}
\caption{(Color online) Event times for the four Twitter users, with the events produced during active states according to the best fitting Lognormal HMM model colored in red (dark grey in greyscale).}
  \label{fig:twostate}
\end{figure}

\begin{figure}[]
  \centering
  \subfloat[]{\label{fig:intro41}\includegraphics[width=0.25\textwidth]{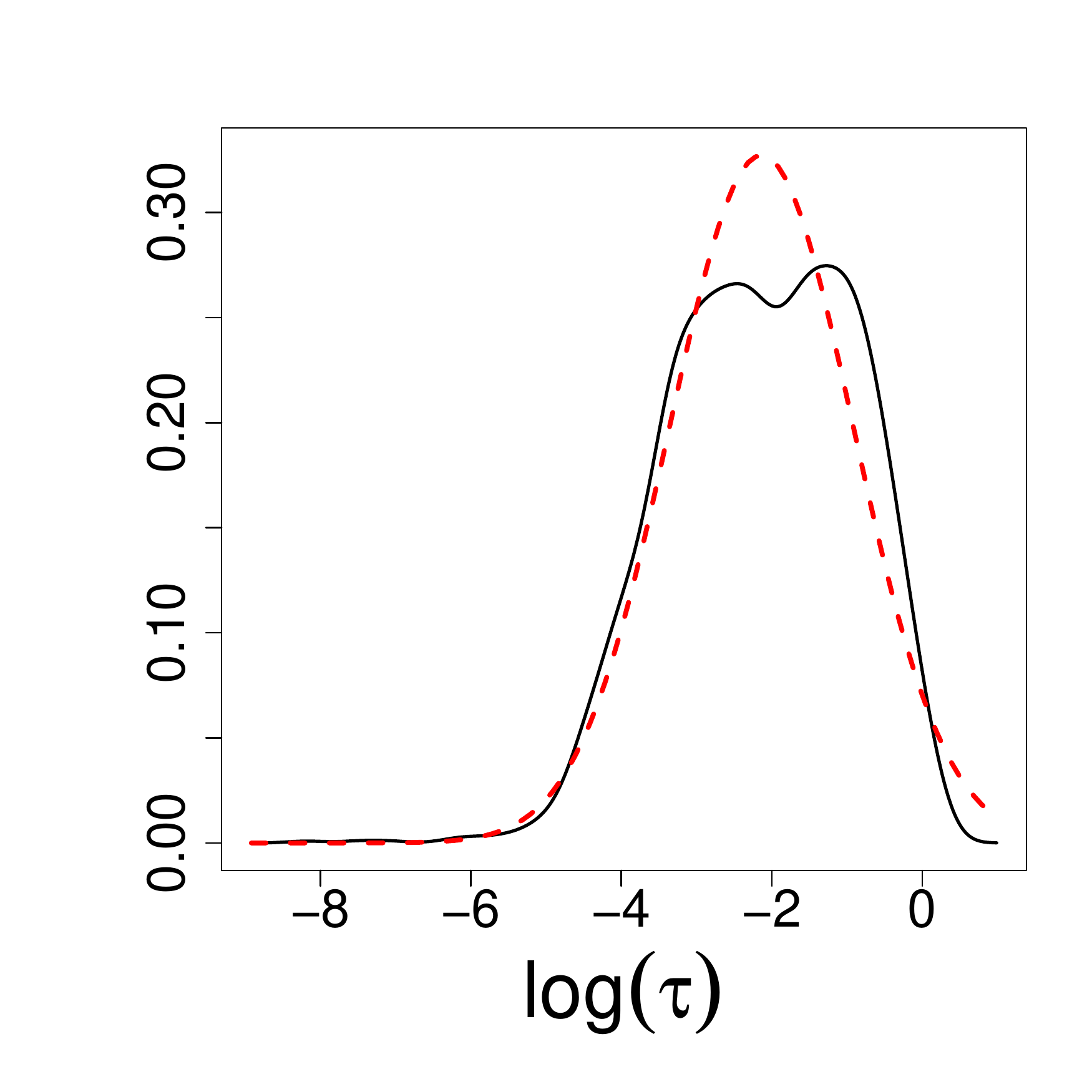}}
  \subfloat[]{\label{fig:intro42}\includegraphics[width=0.25\textwidth]{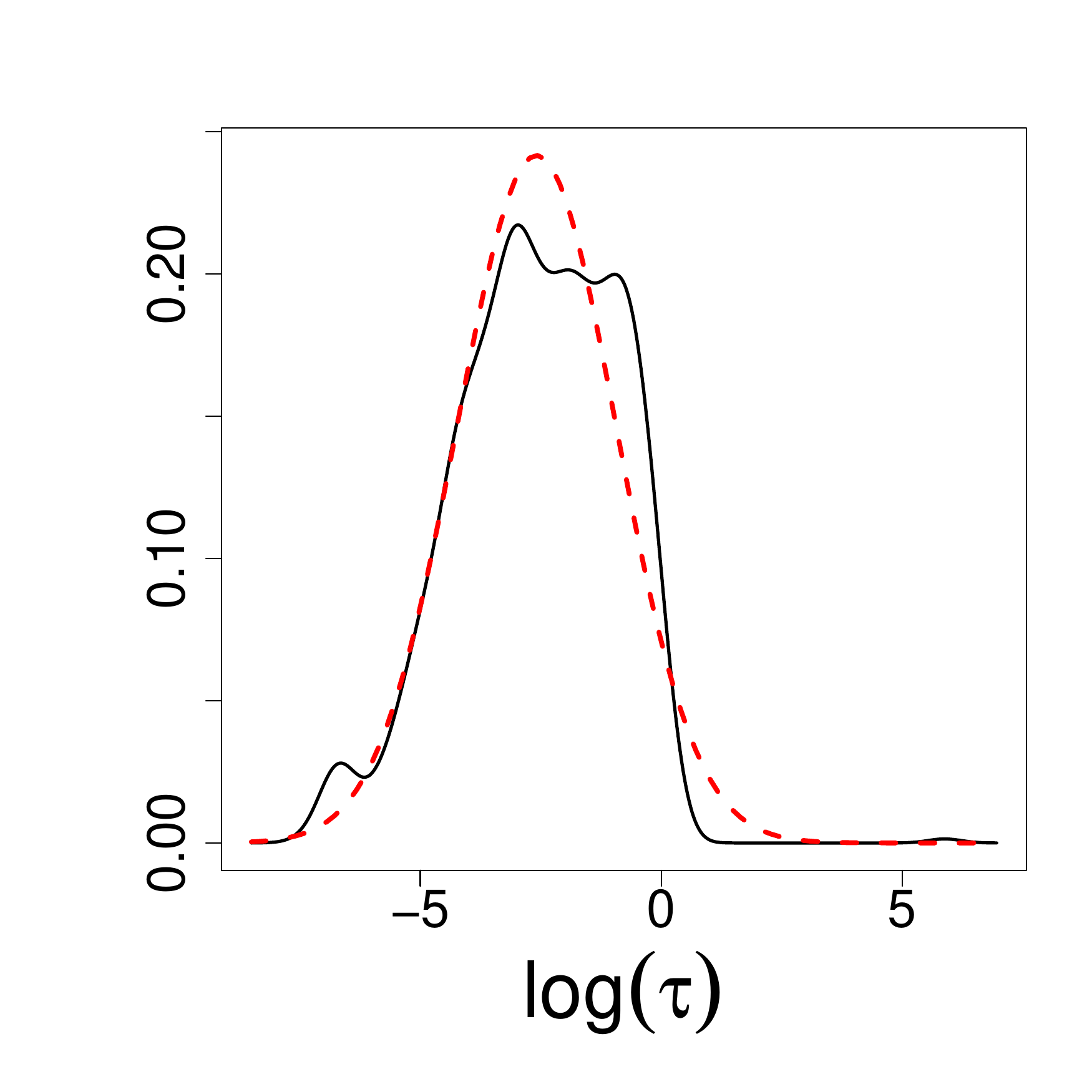}} \\
  \subfloat[]{\label{fig:intro43}\includegraphics[width=0.25\textwidth]{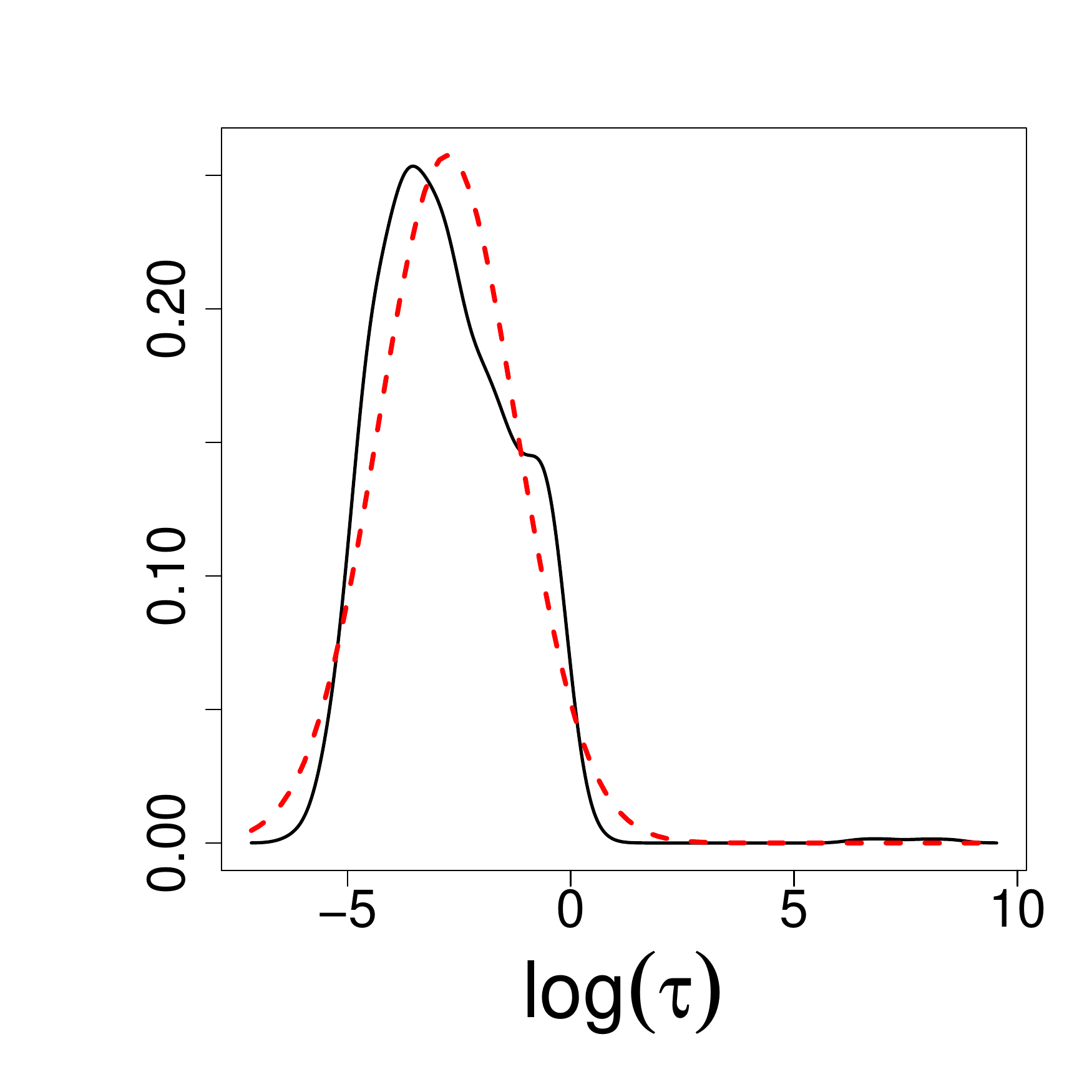}}
  \subfloat[]{\label{fig:intro44}\includegraphics[width=0.25\textwidth]{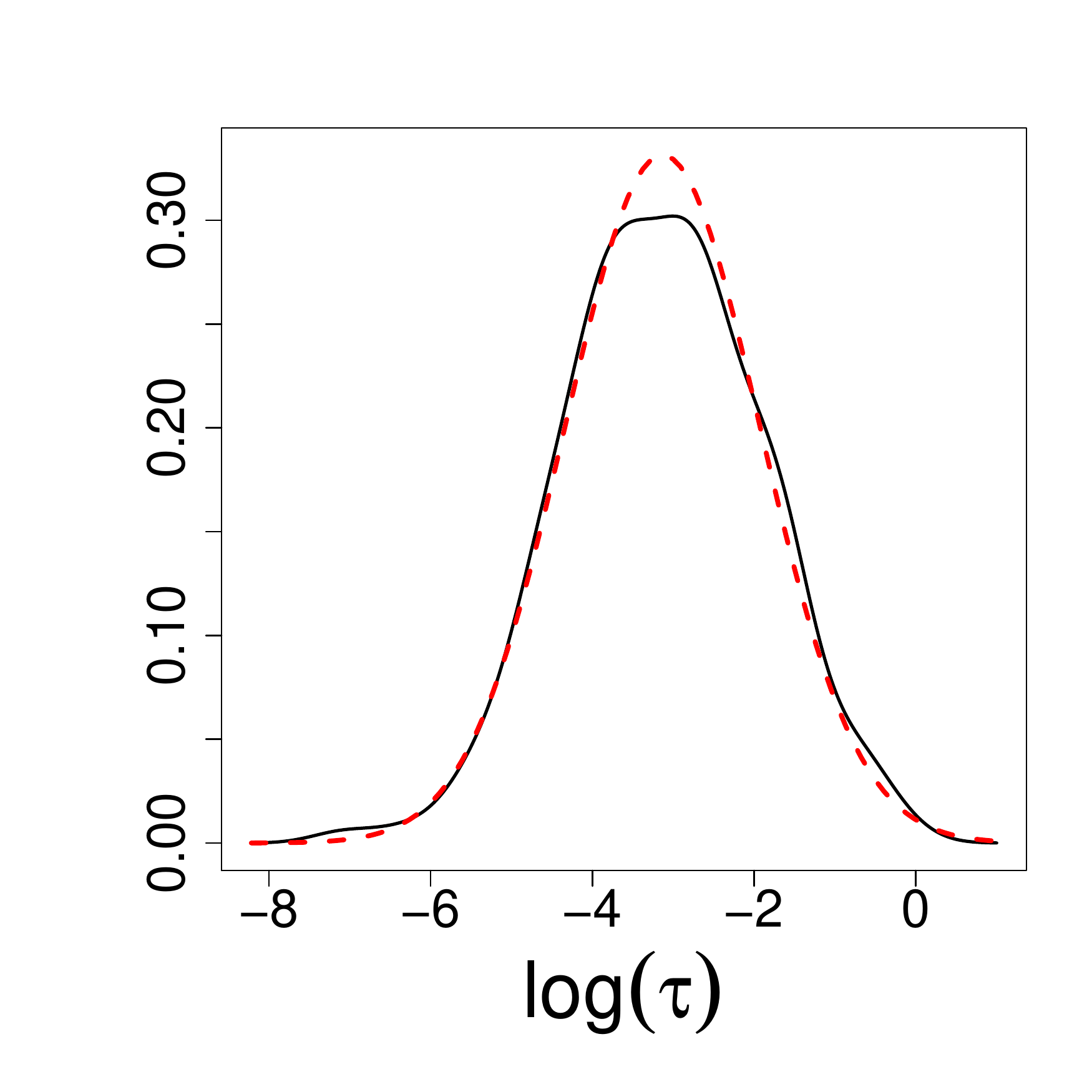}}
\caption{(Color online) Empirical density (black line) of the logarithm of the active state inter-event times for each of the four users, after fitting the HMM using the Lognormal distribution. The dotted red line denotes the best fitting lognormal distribution (note: the logarithmic transformation has been used to aid interpretation)}
  \label{fig:twostatedensity3}
\end{figure}

Table I shows the penalized likelihoods for the four Twitters users previously shown in Fig. 1. It can be seen that for each user, the model using the Lognormal distribution has a substantially higher likelihood than the other two models, indicating that it gives the best fit. The improvement from using the Lognormal model compared to the Exponential model is substantial, which is not surprising given the poor Exponential fit previously noted in Fig. \ref{fig:twostatedensity}. As additional visual evidence for the Lognormal model, Fig. \ref{fig:twostatedensity3} shows the empirical active state distributions for the four Twitter users after fitting the HMM with a Lognormal active state distribution. This clearly gives a better fit than that shown in Fig. \ref{fig:twostatedensity2} when the Exponential distribution was used. Note that the Weibull distribution also constitutes an improvement over the Exponential in all cases, despite being inferior to the Lognormal.

Since considering only four users is not enough to form a definitive conclusion, we applied the same analysis to each of the 10,000 Twitter users in our sample. The results are shown in Table II and it can be seen that in most cases, the Exponential model was overwhelmingly rejected, with it giving the best fit to only $13\%$ of the users. The Lognormal model was clearly the superior model, being chosen as the best fit for $51\%$ of the users, while the Weibull was the best fit for $36\%$. As such, we can conclusively reject the SPB model for social media behavior, with the consequence that human behavior does not seem to be Poissonian, even after accounting for circadian rhythms and burstiness. While this is perhaps disappointing from an aesthetic perspective since it suggests that heavy-tailed behavior is not just an emergent phenomena that results from locally light-tailed Poisson bursts, the basic insights underlying the SPB -- that human behavior is fundamentally dynamic and incorporates both circadian rhythms and burstiness -- do appear to be correct. This can be seen from Fig. 4 which shows the classification of tweets into inactive and active states under the Lognormal model for the four individuals considered throughout. The red dots correspond to events produced during the active state, and it can clearly be seen that these do correspond to genuine bursts in the data. The HMM with Lognormal active state inter-event times appears to describe human behavior very well and would hence be appropriate in any of the myriad of circumstances where it is important to have an accurate model for behavior \cite{Neil2014, Scott2002, Iribarren2009}.


\vspace{-3mm}

\begin{table}[ht]
\centering
\begin{tabular*}{\hsize}{@{\extracolsep{\fill}}rrrr}
  \hline
 Individual & Exponential & Lognormal & Weibull\\ 
  \hline
1 & -2883 & -2832 & -2875\\ 
  2 & -3381 & -3213 & -3227\\ 
  3 & -3114 & -2874 & -3044\\ 
  4 & -1076 & -883 & -1071\\ 
   \hline
\end{tabular*}
\caption{AIC-penalized logliklihoods for each of the three inter-event time distributions for the four considered users. Higher (`less negative') values correspond to a better fit}
\label{tab:twostate}
\end{table}

\vspace{-3mm}

  \begin{table*}[ht]
\centering
\begin{tabular*}{\hsize}{@{\extracolsep{\fill}}rrrr}
  \hline
& Exponential & Lognormal & Weibull\\ 
  \hline
Mean AIC & -722.4 &-703.8 &-695.2\\
Median AIC & --638.6 &-616.5 &-624.8\\
\hline
Best Fit & 0.13 & 0.51 & 0.36 \\

   \hline
\end{tabular*}
\caption{Mean and median AIC-penalized logliklihoods for each of the three inter-event time distributions over the 10,000 Twitter users. The third row shows the proportion of times each distribution was selected as the best fitting model}
\label{tab:results1}
\end{table*}


\section{Evidence for Multiple Types of Human Activity}

Previously we noted that the communication patterns of humans on Twitter are interesting because they are likely to emerge from two distinct types of behavior. The first type of behavior represents  conversations between two Twitter users, and is similar in nature to other conversational behavior such as email and SMS messaging. The second type of behavior represents broadcast communication, where users share messages with all their followers, and these messages are not necessarily part of any particular conversation. This latter behavior differs from that typically studied in the human communication literature, since it is asynchronous rather than responsive.

This observation suggests that the two-state model discussed in the previous section, where users are either in the inactive or active state may be over simplistic. Specifically, we might expect to find two or more distinct types of active states, which correspond to to the two types of behavior noted above. Note that the previous plots of the active state inter-event distributions in Fig. 3 also suggests evidence of multiple types of active states, since they appear to have some degree of bimodality. This bimodality is what would be expected if two distinct types of active behavior were collapsed into a single type, as  noted in a different context by \cite{Wu2010}.

An intuitively plausible qualitative model is the following: suppose that for large parts of the day an individual is inactive, as in the previously discussed models. When the person becomes active, we now suppose that there are two distinct types of states they can transition into, which we will denote by active$^1$ and active$^2$. In the active$^1$ state, the person is engaged in a conversation with one or more other Twitter users, and we would expect to see bursts of events with very short inter-event times, as in SMS messaging \cite{Wu2010}. In the active$^2$ state, the person is present at a device which they can use to post messages, but are not engaged in any specific conversations. Their behavior in this state consists of sending broadcast messages to their followers. As such, we would expect longer inter-event times in this state compared to the conversational active$^1$ state, yet these should still be much shorter than the inter-event times in the inactive state.

We now create a quantitative model which incorporates this hypothesis so that it can be tested. This model is based on the heavy-tailed HMM from the previous section, except that the state space is now enlarged to allow for multiple types of active states. 
As before, we assume that the majority of time is spent in the inactive state, and that transitions into an active state are modeled by an inhomogeneous Poisson process with intensity function $\lambda_0(t)$. When such a transition occurs, the type of active state is randomly selected, with the active$^1$ state (which we will call state 1) selected with probability $p_1$, and the active$^2$ state (which we will call state 2) selected with probability $p_2$. When in state 1, the number of events generated has a Geometric distribution with parameter $\theta^1$, and the inter event times have distribution $g(\tau_i | \gamma^1)$, where again we consider Exponential, Lognormal, and Weibull specifications for this distribution. Similarly, when in state 2, the number of events has a Geometric distribution with parameter $\theta^2$, and the inter-event distribution is $g(\tau_i | \gamma^2)$. In summary, the model for the inter-event times is:


$$p(\tau_i | s_i, t_i, \Theta) \sim  \left\{ \begin{array}{ll}
 \mathrm{Exponential}(\tau_i | \Lambda_{t_i}^{t_i+\tau_i}) &\mbox{if $s_i=0$} \\ 
  g(\tau_i | \gamma^1) &\mbox{if $s_i=1$,}\\
  g(\tau_i | \gamma^2) &\mbox{if $s_i=2$}
       \end{array} \right.
$$
where $0$ denotes the inactive state as before. The parameter vector associated with each user is now $\Theta = (\lambda_0(\cdot), p_0, p_1, p_2, \theta^1, \theta^2, \gamma^1, \gamma^2)$ and this can again be estimated from the data using the standard Baum-Welch algorithm, with details provided in the Appendix.


We compare these models using three HMM states (with two types of active states) to the previous models which used only two HMM states (with one active state). As before, this is done by using the log likelihood, penalized by the AIC. The number of parameters ($k$) is now $9$ when using the Exponential distribution for $g(\cdot)$, and $11$ when using the Lognormal and Weibull distributions. Table 2 shows the resulting penalized likelihoods. Compared to the previous Table I for the models which only used a single type of active state, it is clear that these new models represent a substantial improvement regardless of the functional form used for $g(\cdot)$. This suggests that our hypothesis that there are multiple types of user behavior on social media is correct, and that a model which uses only a binary active/inactive dichotomy is an oversimplification. As before, the model with the Lognormal active state distribution appears to be slightly better than the Weibull.

\begin{table}[h]
\centering
\begin{tabular*}{\hsize}{@{\extracolsep{\fill}}rrrr}
  \hline
 Individual & Exponential3 & Lognormal3 & Weibull3\\ 
  \hline
1 & -2845 & -2825 & -2837\\ 
  2 & -3253 & -3175 & -3189\\ 
  3 & -3024 & -2684 & -2822\\ 
  4 & -1043 & -851 & -987\\ 
   \hline
\end{tabular*}
\caption{AIC-penalized logliklihoods for each of the models which use three states (two active). Higher (`less negative') values correspond to a better fit}
\label{tab:threestate}
\end{table}

To test this more thoroughly, we applied this analysis to all Twitter 10,000 users in our data set. Since the Weibull distribution takes substantially longer than the other models to fit due to the lack of a closed form maximum likelihood estimate for the  model parameters, we omitted it and focused only on the Exponential and Lognormal models since our primary goal here is to compare the three state models to the two state models rather than comparing difference choices of $g(\cdot)$ as in the last section. Our results are shown in Table IV where it can be seen that the three state model fits better for $73\%$ of users ($54\%$ of the time with a Lognormal active state distribution, and $19\%$ of the time with an Exponential distribution).We found that the three state models are more likely to be chosen for users which have a larger number of tweets which is to be expected since more active users are more likely to have a larger number of active state events which helps outweigh the constant penalization enforced by the AIC.  Table IV also reports the mean and median AIC over all 10,000 users, again showing an advantage for the three state model.

  \begin{table*}[ht]
\centering
\begin{tabular*}{\hsize}{@{\extracolsep{\fill}}rrrrrr}
  \hline
& Exponential2 & Lognormal2 & Weibull2 & Exponential3 & Lognormal3\\ 
\hline
Mean AIC & -708.1 &-702.1& -691.3& -675.6 &-661.5\\
Median AIC & -631.6 &-611.5 &-619.5& -606.7 &-599.9  \\
\hline
Best Fit & 0.08 & 0.06 & 0.13 & 0.19 & 0.54\\
   \hline
\end{tabular*}
\caption{Mean and median AIC-penalized logliklihoods for each of the inter-event time distributions over the 10,000 Twitter users, using both two HMM states (one active), and three HMM states (two active). The third row shows the proportion of times each distribution was selected as the best fitting model}
\label{tab:results2}
\end{table*}

As such, we can conclude that at least in the case of social media, human behavior seems to arise due to interactions between multiple types of active states, but that even after accounting for this, there still remains residual heavy tailed behavior which is not well-captured by assuming Exponential inter-event times, and so the assumption of locally Poisson behavior is not viable. The fact that the Lognormal distribution seems to provide the best fit for most users, combined with the central limit theorem, suggests that this more fundamental heavy tail character may be a result of multiplicative interactions between the myriad of factors which govern how people behave. 






\vspace{-3mm}

\section{Discussion}

Most approaches to modelling human communication behavior fall into two camps. The first holds that the heavy tails observed in most empirical records are fundamental aspects of human behavior, such as the hypothesized priority queueing mechanisms that individuals use to schedule their tasks. The second holds that these heavy tails are only present in aggregated data due to averaging over locally Poissonian behavior, mediated by circadian rhythms and burstiness. 

Our results suggest that models from the second class are not sufficient to describe certain types of human activity and that while they partially explain the heavy tails present in aggregate data, there are residual heavy tails which remain unexplained. This suggests that there are more fundamental mechanisms at play. Based on our data analysis, we introduced a new type of quantitative model which allows for circadian rhythms and burstiness as in \cite{Malmgren2008,Malmgren2009}, but which also explicitly incorporates a heavy tailed inter-event time distribution. This was shown to provide a substantially better fit to the social media data considered, again suggesting that some degree of heavy-tailed behavior is fundamental.

Finally, we showed that it is important to avoid oversimplifications which assume that there is only a single type of active behavior, as is often done in the literature. At least when it comes to social media, there seems to be multiple ways in which people behave, corresponding to both conversations with others, and to broadcast behavior which is not typically part of a conversation. Our ultimately conclusion is that human behavior is perhaps less uniform than previously suspected, but that it can still be well captured by quantitative models, provided that they are sufficiently flexible.
  
  \vspace{2mm}
  
  .

  \begin{appendix}

\section{Model Fitting}

\paragraph{Parameterization}: Although there are several possible methods for parametrizing the inhomogenous intensity function $\lambda_0(\cdot)$ which governs the transition from inactive to active states, we chose a simple discretization which represents $\lambda_0(\cdot)$ as a step function. Specifically, we broke the day down into 24 bins each one-hour long, with the function assumed to be constant within each bin. This means that 24 parameters are required to fit $\lambda_0(\cdot)$, corresponding to the value in each bin, and these were estimated during the Baum-Welch phase of the fitting algorithm (see below). Note that this parameterization assumes that the Twitter behavior of a particular user is roughly homogenous across different days of the week, with the behavior on a Monday being identical to that on a Sunday. This seems intuitively unrealistic, so we also considered a paramaeterization where $\lambda_0(\cdot)$ differed on each day of the week, corresponding to $7 \times 24 =  168$ parameters, and a  paramaeterization which assumed that all weekdays were homogenous but where a different function was fit to weekends, which corresponds to $2 \times 24 = 48$ parameters. Neither of these modifications change the conclusions found in the main text.

\paragraph{Fitting the Hidden Markov Model}: If a HMM can be written in standard discrete time form then the parameters can be estimated using a combination of the Baum-Welch and Viterbi algorithms, will full details provided in a standard reference such as \cite{Rabiner1989}. This can be done for our model using the emission distributions discussed in the main text, and the transition matrix:

\vspace{-3mm}

$$
Q =  \left( \begin{array}{ccc}
1-p & p\\
1-\theta & \theta
\end{array}  \right)
$$

\vspace{-3mm}

The state-specific emission distributions are given in the text for both the SPB model and our various extensions. This formulation allows the Baum-Welch and Viterbi algorithms to be implemented directly.  The extension to the models with multiple active states is direct, and parallels estimation for multiple state Hidden Markov Models. Note that we forbid direct transitions from one type of active state to another unless the inactive state is passed through first, in order to make each state distinct and aid interpretability. Allowing direct transitions would produce `blurred' active bursts which pass through several active states and have no clear interpretation.

  \end{appendix}

\bibliographystyle{apsrev4-1}




\end{document}